# Electron beam irradiation effects on the structure and mechanical properties of PEEK. Part 1.

N. Almassov[1], B. Rakhadilov[2], N. Serik[1], Zh. Alsar[1], Zh. Sagdoldina[2], G. Andybayeva[3], N. Zhakiyev[4], C. Spitas[1], C. Kostas[1], Z. Insepov[1,5,6] *

[1] Nazarbayev University, Nur-Sultan 010000, Kazakhstan
[2] Amanzholov University, 070020 Ust-Kamenogorsk, Kazakhstan
[3] D. Serikbayev East Kazakhstan Technical University
[4] Astana IT University, Nur-Sultan 010000, Kazakhstan
[5] National Nuclear Research University (MEPhI), Moscow 115409, Russian Federation
[6] Purdue University, West-Lafayette, IN 47907, USA
*Corresponding author email: zinsepov@purdue.edu

**Abstract**

Effects of electron-beam processing on the structure and mechanical properties of poly-ether-ether ketone (PEEK) were studied by XRD and tribological tests. An ambiguous change in the mechanical properties of the irradiated samples was found. In particular, the wear rate of sample 1 is higher than that of the unirradiated original sample, while it is lower for samples 2-7. Sample 4 had the lowest wear rate, and most likely the changes in the mechanical properties of the samples were due to the transformation of their amorphous component. Comparison of the X-ray spectra of the irradiated and unirradiated samples did not reveal any differences in their crystal structure.

## 1 Introduction

Polyether ether ketone (PEEK) is a semi-crystalline thermoplastic with excellent thermomechanical properties that are resistant to chemically active and radiation environments [1]. Thanks to these properties, the polymer is widely used in the automotive [2], aerospace [3], nuclear [4], tribological [5] and other industries. However, there has been limited research to understand the effect of the molecular structure of PEEK on the properties of bulk materials. The improvement of mechanical, thermal and tribological properties under the influence of irradiation has been reported in a number of works. This article examines the effect of electron beam irradiation on the structure and tribological properties of PEEK.

## 2 Materials and methods

*PEEK samples*
Polyetheretherketone (PEEK) was purchased from Ensinger. This polymer is a high performance, high temperature, semi crystalline thermoplastic manufactured by Ensinger and may use Victrex® PEEK 450G or Solvay's KetaSpire® KT-820 polymer.

*Electron-beam processing*
Electron-beam processing or electron irradiation (EBI) of poly-ether-ether-ketone (PEEK) was carried out on an industrial pulsed accelerator ILU-10 at the Park of Nuclear Technologies JSC (Kurchatov, Kazakhstan). Taking into account the cumulative effect of EBI on the properties of polymers, the irradiation regimes were associated with a variation in the irradiation dose, which depend on the number of runs, i.e. from the movement of the irradiated materials relative to the electron beam on a moving table.

*Tribological tests*
Tribological tests of PEEK polymers were carried out before and after EBI. The friction-sliding tribological test was performed on a TRB3 tribometer using the standard ball-and-disk technique. A ball with a diameter of 6.0 mm made of steel - 100Cr6 was used as a counterbody. The tests were carried out at a load of 10 N and a linear velocity of 5 cm / s, a radius of curvature of wear of 2 mm.
The tribological characteristics of the samples before and after EBI were characterized by the rate of wear. Tip wear rate is calculated based on the volume of material displaced during the test. The volume of polymer wear after the tribological test was determined using a Profiler 130 (the profile of the wear track was measured).

*XRD*
To assess the effect of E-beam processing on the structural-phase state of PEEK, X-ray phase analysis was carried out on an X'Pert PRO diffractometer (PANalytical, Netherlands). Shooting modes: diffraction angle from 10 ° to 45 °; scanning step 0.03; exposure time 0.5 s; radiation: Cu Kα; voltage and current: 45 kV and 30 mA.

## 3 Results and discussion
The modes of EBI of PEEK are presented in Table 1.
Table 1 – EBI modes of PEEK samples on the ILU-10 accelerator

| Samples | Beam energy, MeV | Beam curent, mA | Speed, m/min | Number of runs | Radiation dose for 1 run, kGy | Total radiation dose, kGy |
|---|---|---|---|---|---|---|
| PEEK-1 | 2,7 | 6,84 | 9 | 5 | 10 | 50 |
| PEEK-2 | 2,7 | 6,84 | 9 | 10 | 10 | 100 |
| PEEK-3 | 2,7 | 6,84 | 9 | 20 | 10 | 200 |
| PEEK-4 | 2,7 | 6,84 | 0,8 | 2 | 100 | 200 |
| PEEK-5 | 2,7 | 6,84 | 9 | 30 | 10 | 300 |
| PEEK-6 | 2,7 | 6,84 | 0,8 | 3 | 100 | 300 |
| PEEK-7 | 2,7 | 6,84 | 9 | 40 | 10 | 400 |

The friction coefficient of PEEK samples before and after EBI are shown on Figure 1.

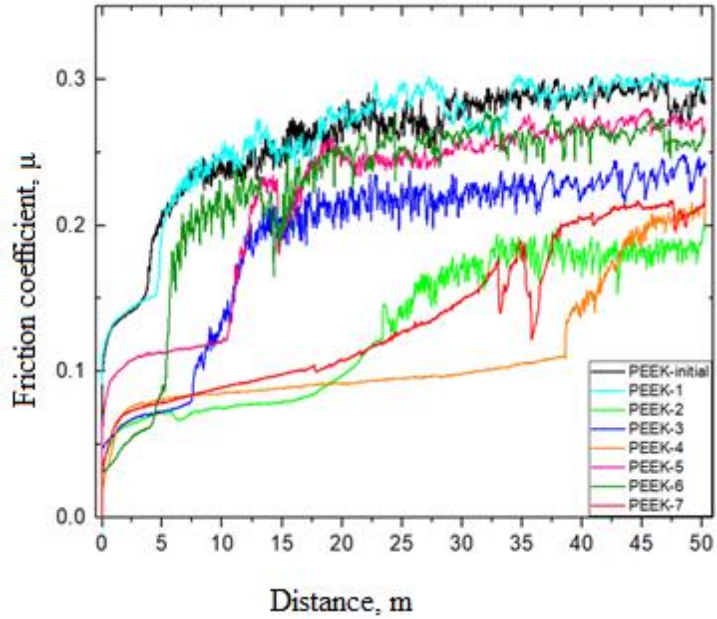

Figure 1. Graphs of the friction coefficient μ of PEEK samples

The wear rate under the action of the tip is calculated based on the volume of material displaced during the test, which was calculated using the following formula (1):

$$I = \frac{V}{F \times l} \qquad (1)$$

where,

I - intensity of wear, [mm$^3$ / N • m]; l - friction path, [m]; F - nominal pressure, [H]; V is the volume of the worn out part, [mm3].

Figure 2 shows the result of calculating the PEEK wear rate before and after ELO. The results of the study showed a high value of the wear rate of PEEK-1 (radiation dose 50 kGy) in comparison with the original sample.

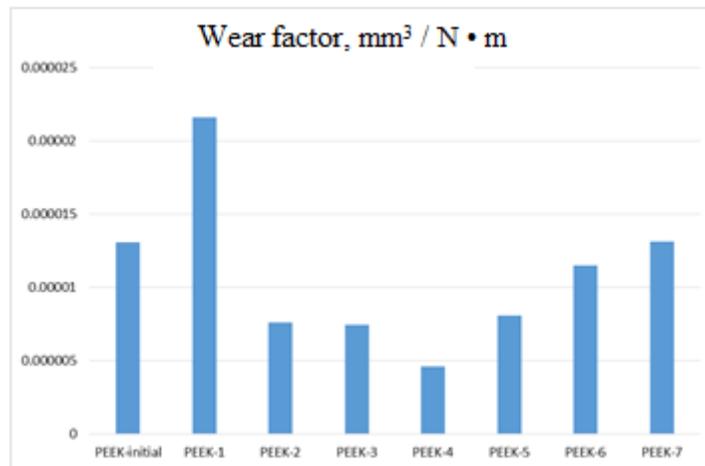

Figure 2. Wear rate of PEEK samples

Figure 3 shows diffractograms of PEEK before and after EBI. There are four diffraction peaks with maximum angles of 18.717 °; 20.709 °; 22.677 ° and 28.732 °. The diffraction patterns at different irradiation doses are practically the same, which indicates the minimal effect of ELO on the crystal structure of PEEK. There are slight differences in the height and width of the diffraction peaks at different irradiation doses, which suggests a change in the degree of crystallinity (ratio of amorphous to crystalline phases) of PEEK. The crystallinity has a great influence on the tribological and mechanical characteristics of polymers.

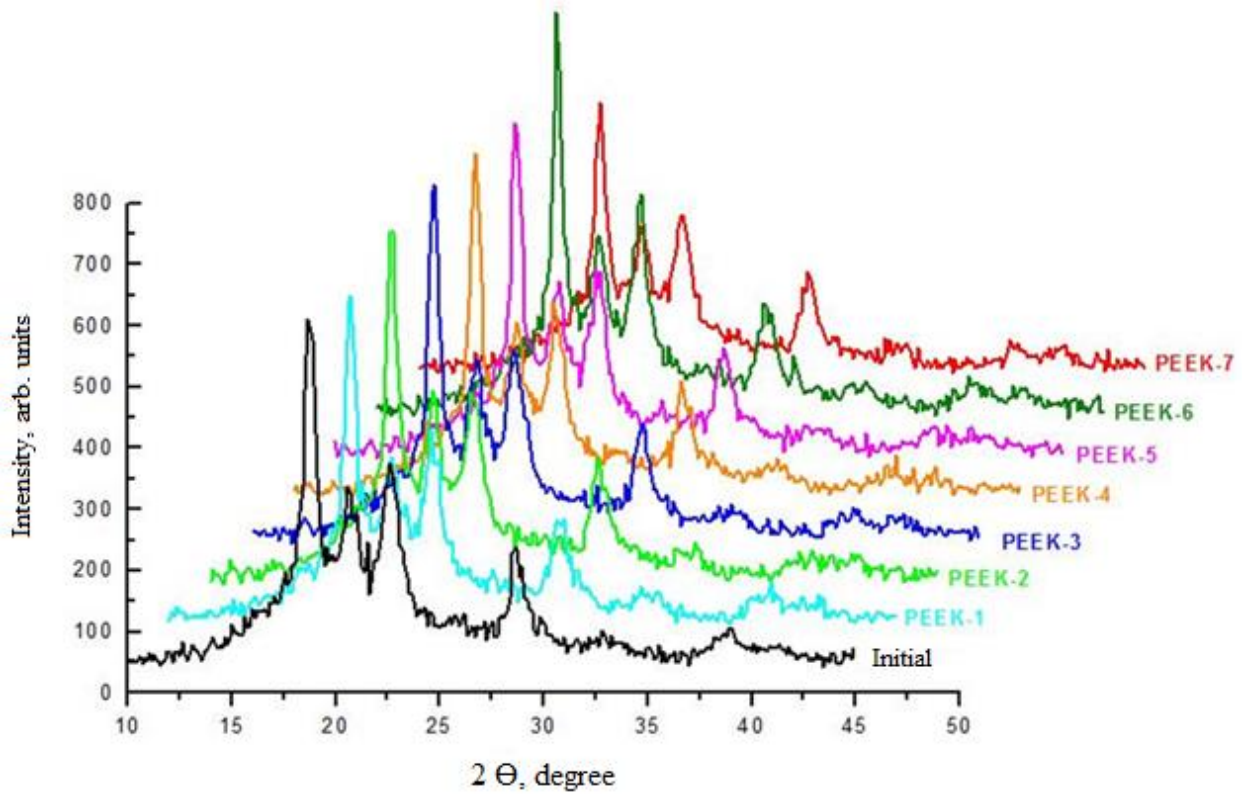

Figure 3. XRD pattern of PEEK samples

An analysis of the XRD-spectra of the initial unirradiated sample and the sample with the greatest effect of exposure to radiation (PEEK 4) is shown in Figure 4, 5 respectively. The decoded sample parameters are shown in Table 2, 3.

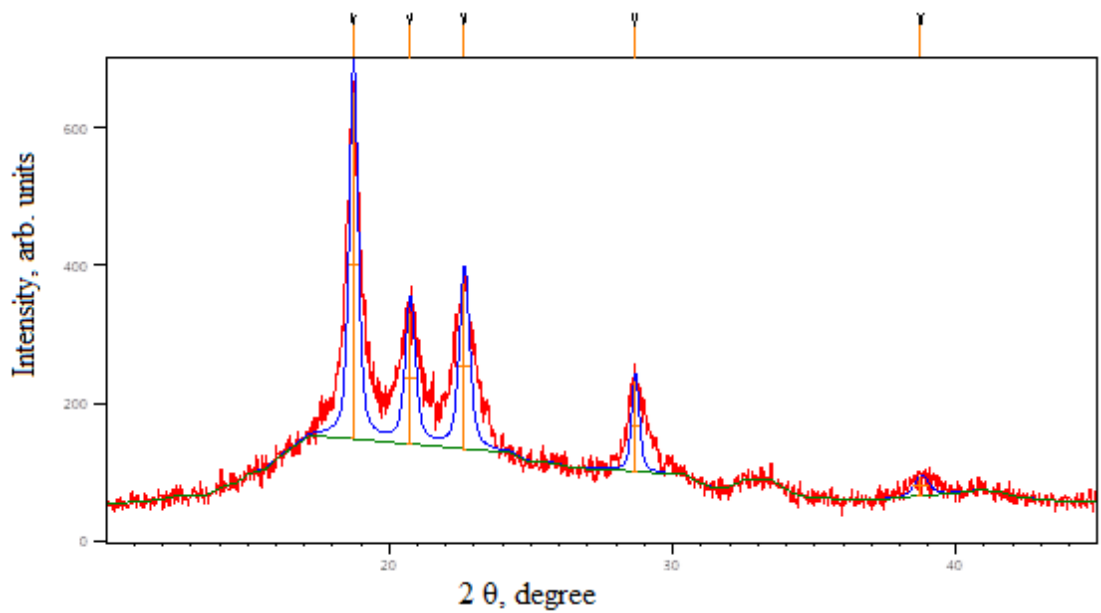

Figure 4. An analyze view of XRD-spectra of initial unirradiated PEEK sample

| 2θ [°] | Intensity [arb. units]] | FWHM [°2θ] | Interlayer distance [Å] | Rel. int. [%] |
|---|---|---|---|---|
| 18,7316 | 504,15 | 0,3542 | 4,73732 | 100,00 |
| 20,7309 | 191,75 | 0,4133 | 4,28474 | 38,03 |
| 22,6433 | 240,63 | 0,4133 | 3,92701 | 47,73 |
| 28,6634 | 133,99 | 0,2952 | 3,11445 | 26,58 |
| 38,7528 | 27,13 | 0,5904 | 2,32369 | 5,38 |

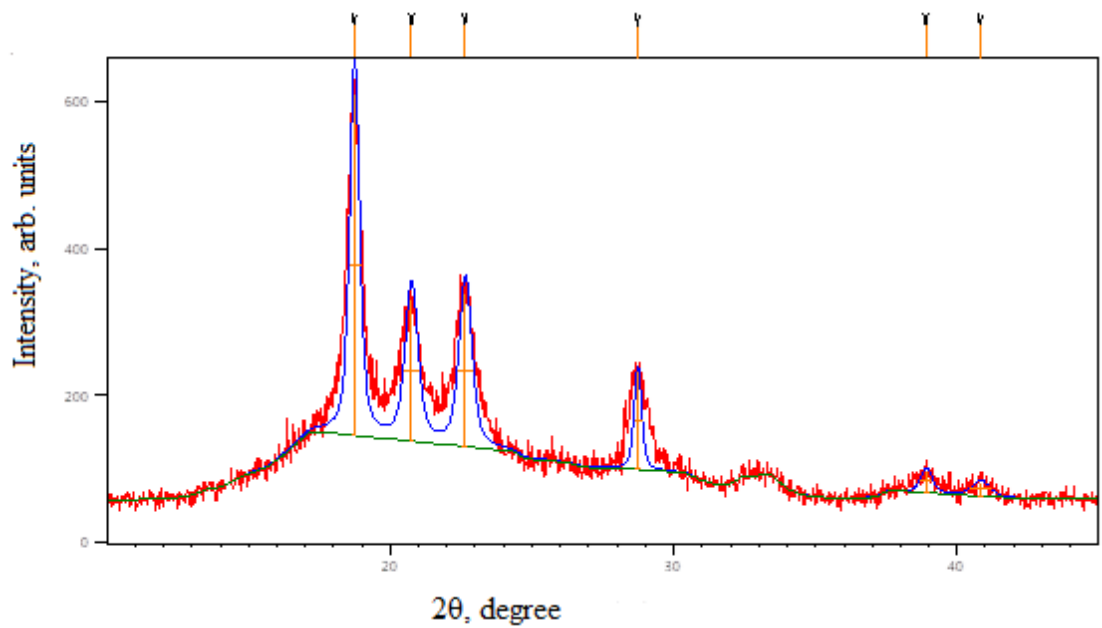

| 2θ [°] | Intensity [arb. units] | FWHM [°2θ] | Interlayer distance [Å] | Rel. int. [%] |
|---|---|---|---|---|
| 18,7301 | 465,59 | 0,4133 | 4,73768 | 100,00 |
| 20,7391 | 193,41 | 0,5314 | 4,28307 | 41,54 |
| 22,6420 | 208,16 | 0,5314 | 3,92723 | 44,71 |
| 28,7134 | 133,54 | 0,2952 | 3,10914 | 28,68 |
| 38,8886 | 30,72 | 0,4723 | 2,31589 | 6,60 |
| 40,8134 | 19,86 | 0,7085 | 2,21101 | 4,27 |

## 4 Conclusions

The improvement in the mechanical properties of PEEK polymer exposed to electron irradiation has been established using tribological tests. A preliminary analysis of the irradiated samples using X-ray spectroscopy did not reveal significant changes in their atomic structure.

**AKNOWLEGMENTS**

This work has been funded by the Nazarbayev University Collaborative Research Project (CRP): "Development of smart passive-active multiscale composite structure for earth Remote Sensing Satellites (RSS) of ultrahigh resolution (ULTRASAT)", Grant Award Nr. 091019CRP2115.